\documentclass[prd,superscriptaddress,nofootinbib,preprint]{revtex4-1}
\usepackage{graphicx}
\usepackage{dcolumn}
\usepackage{bm}
\usepackage{amssymb}
\usepackage{mathrsfs}
\usepackage{amsmath}
\usepackage{epsfig}
\usepackage[dvips]{color}
\usepackage{hhline}
\usepackage[colorlinks]{hyperref}
\usepackage[normalem]{ulem}
\usepackage[dvips]{color}
\usepackage{hhline} 
\usepackage[colorlinks]{hyperref}
\newcommand{\be}{\begin{equation}}
\newcommand{\ee}{\end{equation}}
\newcommand{\bea}{\begin{eqnarray}}
\newcommand{\eea}{\end{eqnarray}}
\newcommand{\ca}{\cal A}
\begin{document}
\title{Can the observed CP asymmetry in  $\tau \rightarrow K\pi\nu_{\tau}$ be due to Non-Standard Tensor interactions?}
\author{H. Zeen Devi}
\email{zeen@imsc.res.in}
\affiliation{Institute of Mathematical Sciences,Chennai, India}
\author{L Dhargyal}
\email{dhargyal@imsc.res.in}
\affiliation{Institute of Mathematical Sciences,Chennai, India}
\author{Nita Sinha}
\email{nita@imsc.res.in}
\affiliation{Institute of Mathematical Sciences,Chennai, India}
\begin{abstract}
An intriguing opposite sign of the CP violating asymmetry was recently measured in the tau decay modes  $\tau^{\pm} \rightarrow K_s \pi^{\pm} \overset{(-)}{\nu_{\tau}} $ by the BaBar collaboration, than that expected within the Standard Model. If this result is confirmed with higher precision, the observed decay rate asymmetry $A_{CP}$, can only arise from some Non-Standard Interactions(NSI) occurring possibly in both the hadronic as well as in the leptonic sectors. We illustrate that while a simple charged scalar interaction cannot yield this rate asymmetry, it will be possible to generate this in the presence of a tensor interaction. Parameterizing the strength and weak phase of this  NSI contribution, the observed branching ratio and the decay rate CP asymmetry for the particular mode $\tau^{\pm} \rightarrow K_s \pi^{\pm} \overset{(-)}{ \nu_{\tau}} $, are used to determine the CP violating weak phase and the coupling of a tensorial interaction that can give a consistent sign and magnitude of the asymmetry. 
\end{abstract}
\maketitle
\section{Introduction}
While the Kobayashi Maskawa ansatz for CP violation within the Standard Model(SM)~\cite{Kobayashi:1973fv} in the quark sector has been clearly verified by the plethora of data from the B factories, this is unable to account for the observed Baryon Asymmetry of the Universe (BAU). Hence one needs to look for other sources of CP violation, including searches in the leptonic sector. Apart from the CP phases that may arise in the neutrino mixing matrix, the decays of the charged tau leptons may allow us to explore non-standard CP violating interactions.
   
In fact, CP violation through decays of tau leptons had been studied in a series of papers by Tsai~\cite{Tsai:1994rc}, around the time of the preliminary design for a tau-charm factory. Various experimental groups have been involved in exploring CP violation in tau decays in the last decade or more.
In 2002, the CLEO collaboration \cite{Bonvicini:2001xz} and more recently the Belle Collaboration \cite{Bischofberger:2011pw} studied the angular distribution of the decay products in  $\tau^{\pm}  \rightarrow K_s^0 \pi^{\pm} \overset{(-)}{\nu}_{\tau}$
in search of  CP violation, however, neither study revealed any CP  asymmetry.
The BaBar collaboration \cite{BABAR:2011aa} for the first time reported a measurement of the rate asymmetry  $A_{CP}$ in this decay mode to be: 
\be
A_{CP}^{Exp.}=(-0.36 \pm 0.23 \pm 0.11)\%.
\ee
On the theoretical side, for $\tau^{\pm}  \rightarrow K_s^0 \pi^{\pm} \overset{(-)}{\nu}_{\tau} \rightarrow {[\pi^+ \pi^-]}_{K} \pi^{\pm} \overset{(-)}{\nu }_{\tau}$, Bigi and Sanda \cite{Bigi:2005ts} predicted the CP asymmetry to be,  
\be
 A_{CP}^{SM}=(+ 0.33 \pm 0.01)\%, 
\ee
 where the CP violation arises from the {\it known} $K^0-\bar{K}^0$ mixing. Recently, Grossman and Nir \cite{Grossman:2011zk} comparing the rate asymmetries for decays to neutral kaons, of the taus with that of D mesons, 
pointed out that since $\tau^+(\tau^-)$ decays initially to a $K^0(\bar{K}^0)$ whereas $D^+(D^-)$ decays initially to $\bar{K}^0(K^0)$, the time integrated decay rate CP asymmetry (arising from oscillations of the neutral kaons) of $\tau$ decays must have a sign opposite to that of $D$ decays. Further, they emphasized that the decay asymmetry is affected by the reconstruction efficiency as a function of the $K_s\rightarrow \pi^+ \pi^-$ decay time. Using the parameterization of Ref.~\cite{Grossman:2011zk}, BaBar has obtained a multiplicative correction factor for the decay-rate asymmetry and predicts the SM decay-rate asymmetry to be,
\be
 A_{CP}^{SM}=(+ 0.36\pm 0.01)~\%. 
\ee  
As reported in Refs.~\cite{delAmoSanchez:2011zza,Ko:2010ng,Mendez:2009aa,Link:2001zj} the CP asymmetry in $D^{\pm}\rightarrow\pi K_s$ decay has been measured to be $(-0.54\pm0.14)\%$. 

The observation of a CP asymmetry in tau decays to $K_s$ having the same sign as that in $D$ decays and moreover of the same magnitude but opposite in sign to the corrected SM expectation, implies that this asymmetry cannot be accounted for by the CP violation in $K^0-\bar{K}^0$ mixing. Apart from this mixing contribution to CPV, within the SM, since there is only a single amplitude with the W boson mediating the decay process, the observed CP asymmetry cannot be explained. Hence, this may be a signal of physics beyond the standard model, if confirmed to a higher statistical significance. 

It should be pointed out that the BaBar collaboration has accounted for the modification of the decay rate asymmetry due to different nuclear interaction cross-sections of the  $K^0$ and $\bar{K}^0$ mesons with the detector material through a correction, calculated on an event by event basis.
Also, BaBar~\cite{BABAR:2011aa} claims that using a control sample in data and Monte Carlo simulations, they have verified that no significant decay rate asymmetry is induced by their detector or the selection criteria.  Further, the standard model asymmetry is identical for decays with any number of $\pi^0$ mesons and hence can be searched for, in all the modes $\tau^{\pm} \rightarrow \pi^{\pm} K_s^0(\ge 0\pi^0) \overset{(-)}{\nu }_{\tau}$. We note that all the experiments CLEO, Belle, BaBar assume that the CP asymmetry is conserved at the tau production vertex. 

\begin{figure}[htb]
\begin{minipage}[t]{0.48\textwidth}
\hspace{-0.4cm}
\begin{center}
\includegraphics[width=7cm]{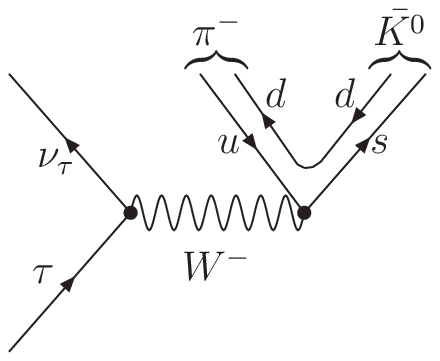}
\end{center}
 \end{minipage}
 \hfill
 \begin{minipage}[t]{0.48\textwidth}
 \hspace{-0.4cm}
 \begin{center}
 \includegraphics[width=7cm]{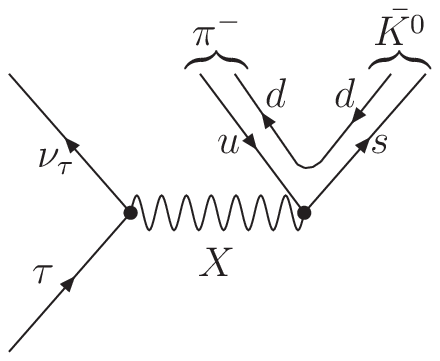}
 \end{center}
 \end{minipage}
 \caption{ The Feynman diagrams for the Standard Model decay mode via W exchange is shown in the left panel and the New physics exchange diagram via some exotic particle X is shown in the right panel.}
\label{feynmann-dia}
 \end{figure}

In the presence of an additional new physics (NP) amplitude with a complex coupling, along with the strong phases from the $K-\pi$ scattering, which can be large particularly in this resonance dominated region, the observed CP asymmetry may be attainable. This Non Standard Interaction (NSI) could possibly affect both the hadronic and leptonic currents.
 The Feynman diagrams for the SM decay mode of tau via W exchange and via exchange of some exotic particle X, in the presence of NP are shown in Figure~(\ref{feynmann-dia}).   

Naively one would expect a charged scalar boson exchange to provide the required additional diagram, where, with a complex weak coupling and the difference in the strong phases of the scalar and vector hadronic form factors, a CP violating asymmetry could arise. However, such an asymmetry appears only in the difference of the $\tau^\pm\rightarrow K_S\pi^\pm\nu_{\tau}(\bar\nu_{\tau})$ decay angular distributions, but vanishes in the integrated difference of the decay rates for $\tau^+$ and $\tau^-$,  measured by BaBar. In fact, CLEO had used their non-observation of an asymmetry in the distribution, to set a bound~\cite{Bonvicini:2001xz} on the imaginary part of the complex coupling of the scalar boson, such as a charged Higgs. Similar limits were set by Belle~\cite{Bischofberger:2011pw} on the CP violation parameter modifying the scalar form factor, since their differential asymmetry was compatible with zero.
The new physics amplitude that can account for a decay rate asymmetry therefore has to appear from a different kind of interaction and we investigate whether a tensor interaction can produce the asymmetry reported by BaBar and in fact, use the measured branching ratio and asymmetry to constrain the parameters of this kind of new interaction.

In section~\ref{RateAsym}, we evaluate the decay rate asymmetry in the presence of any generic NP amplitude. The decay rate for the SM case is calculated in section~\ref{SM-Rate}, while in section~\ref{Total-Rate} the total rate is evaluated in the presence of the new additional tensor interaction term. The observables, the branching ratio and the decay rate asymmetry are used to estimate the parameters of this new interaction in section~\ref{estimation} and in section~\ref{conclusion} we conclude.

\section{Branching Ratio and Rate asymmetry in  $\tau \rightarrow K\pi\nu_{\tau}$  in presence of a generic New Physics amplitude}
\label{RateAsym}
The amplitude for the decay of  $\tau^+ \rightarrow K_s\pi^+\bar{\nu}_{\tau}$ in the presence of NSI  can be written as 
\be
{\cal A}= {\cal A}^{SM} + {\ca}^{NSI}e^{i\phi}e^{i\delta}
\label{decayA}
\ee
where $A^{SM}$ and $A^{NSI}$ are the magnitudes of the SM and NP amplitudes respectively, while $\phi$ is the weak phase of the NP contribution (since $V_{us}$ is real, there is no weak phase in the SM contribution, except that coming from the neutral $K$ meson mixing, which is accounted for, separately in the theoretical SM expectation). The relative strong phase of the $(K\pi)$ system, $\delta$, between the NP amplitude with respect to the SM contribution is a function of the $K\pi$ invariant mass-squared. The amplitude for the anti-process $\tau^- \rightarrow K_s \pi^-\nu_{\tau}$ has the opposite weak phase but the same strong phase.

In presence of a NSI whose interference with SM is non-vanishing even after the angular integrations\footnote{Details about the integration variables are specified in the next section.}, the general expression for the differential decay rate of $\tau \rightarrow K\pi\nu$ may be written as,  
\bea
d\Gamma & \propto & \big|{\cal A}\big|^2\times \, dQ^2,\\
& \propto & \big[ |{\ca}^{SM}|^2 + |{\ca}^{NSI}|^2 + 2|{\ca}^{SM}|\,|{\ca}^{NSI}|\,cos(\phi+\delta(Q^2))\big]\, dQ^2,
\eea
where $Q$ is the sum of the hadron momenta. The differential decay rate for the anti-process is,
\bea
d\bar{\Gamma} & \propto & \big|\bar {\cal A}\big|^2\times \, dQ^2,\\
& \propto & \big[ |{\ca}^{SM}|^2 + |{\ca}^{NSI}|^2 + 2|{\ca}^{SM}|\,|{\ca}^{NSI}|\,cos(-\phi+\delta(Q^2))\big]\, dQ^2.
\eea
The branching ratio for $\tau\rightarrow K_s\pi\nu$ is the ratio of the average of the width of $\tau^+ \rightarrow K_s\pi^+\bar{\nu}$ and 
$\tau^- \rightarrow K_s\pi^-\nu$  to the total width of $\tau$ ($\Gamma_{total}$). Hence,   
\be
BR(\tau\rightarrow K_s\pi\nu) =\frac{\Gamma+\bar{\Gamma}}{2\,\Gamma_{total}}~.
\ee
Now,
\bea
\frac{d\Gamma}{dQ^2}+\frac{d\bar{\Gamma}}{dQ^2} &\propto& 2 \left[  |{\ca}^{SM}|^2 + |{\ca}^{NSI}|^2 + 2|{\ca}^{SM}|\,|{\ca}^{NSI}|\,cos(\phi)\,cos(\delta(Q^2))  \right]~,\\
&\propto & 2\,{\ca}^{SM}|^2 \left[ 1 + (r(Q^2))^2 + 2\,r(Q^2)\,cos(\phi)\,cos(\delta(Q^2)) \right]~,
\eea
where $r(Q^2)$ is the ratio of the amplitude of the NSI contribution to the SM contribution.
Therefore,
\be
BR(\tau\rightarrow K_s\pi\nu) = \frac{\int dQ^2\, 2\,|{\ca}^{SM}|^2 \left[ 1 + (r(Q^2))^2 + 2\,r(Q^2)\,cos(\phi)\,cos(\delta(Q^2)) \right]}{2\,\Gamma_{\tau}}~.
\label{br}
\ee

Similarly the difference,
\bea
\frac{d\Gamma}{dQ^2}\, - \,\frac{d\bar{\Gamma}}{dQ^2} &\propto & -\,4\,|A^{SM}|\, |A^{NSI}|\, sin(\phi) \, sin(\delta(Q^2))~,\\
&\propto& -\,4\, r(Q^2)\,sin(\phi) \,sin(\delta(Q^2))\,\frac{d\Gamma^{SM}}{dQ^2}~.
\eea
 Hence, the integrated rate asymmetry, \\
\bea 
\nonumber
A_{CP}^{\tau}&=&\frac{\Gamma\,-\,\bar{\Gamma}}{\Gamma +
  \bar{\Gamma}}\\ 
&=& -\, \frac{ \int\,4\, r(Q^2)\,sin(\phi) \,sin(\delta(Q^2)) \,
  \frac{d\Gamma^{SM}}{dQ^2} \, dQ^2}{\Gamma + \bar{\Gamma}}~. 
\label{Acp}
\eea
As pointed out in Refs.~\cite{Bigi:2012km} and~\cite{Pich:2009zza}, this CP asymmetry being linear in the new physics amplitude has a higher sensitivity to it, than effects like lepton flavour violation and electric dipole moments etc., which depend quadratically on the NP amplitude, rendering this observation to play an important role in uncovering physics beyond the SM.
Using eqn.(\ref{br}) and eqn.(\ref{Acp})  and the PDG~\cite{PDG-2012} value of the Branching Ratio  and the rate asymmetry measured by BaBar, the weak phase and the magnitude of the new physics contribution can be estimated.

\section{ Decay rate  of   $\tau \rightarrow K \pi \nu $ in the Standard Model}
\label{SM-Rate}

The tau leptonic and the hadronic decay amplitude can be factorized into a purely leptonic part including the tau and the neutrino and a hadronic part, where the hadronic system is created from the QCD vacuum via the charged weak current.

Hence, the differential decay rate of the process $\tau(p_{\tau})\rightarrow K(p_{K})+ \pi(p_{\pi}) + \nu(p_{\nu_{\tau}})$ \cite{Finkemeier:1996dh} may be written as,
\be
d\Gamma(\tau\rightarrow K\pi\nu)= \frac{1}{2m_{\tau}}\frac{G_F^2}{2}sin^2\theta_c{\cal L_{\mu\nu}H^{\mu\nu}}dPS^{(3)}~,
\ee
where ${\cal L_{\mu\nu}}$ is the leptonic term: 
\be
{\cal L_{\mu\nu}}=  [{\bar \nu_{\tau}}\gamma_{\mu}(1-\gamma_5)\tau]\,[{\bar \nu_{\tau}}\gamma_{\nu}(1-\gamma_5)\tau]^{\dag}  
\ee
and the hadronic term: 
\be 
\cal H^{\mu\nu}=J^{\mu}(J^{\nu})^{\dag}~, 
\ee
 is given in terms of the hadronic vector current
\be
{\cal J^{\mu}}= \langle K(p_K)\pi(p_{\pi})|V^{\mu}(0)|0 \rangle.
\label{vc}
\ee
A transition from vacuum to two pseudoscalar mesons can occur only via scalar and vector currents as for the $K\pi$ system the allowed values of $J^P$ are $0^+$ and $1^-$. The hadronic vector current in eqn.(\ref{vc}) is parameterized in terms of the scalar and the vector form factors as,
\be
{\cal J^{\mu}}=F_V^{K\pi}(Q^2)\big( g^{\mu\nu} - \frac {Q^{\mu}Q^{\nu}}{Q^2}\big) (p_k-p_{\pi})_{\nu}\,+ F_s^{K\pi}Q^{\mu},
\ee
where $Q^{\mu}=p_{k}^{\mu} +p_{\pi}^{\mu}$.
The decay rate involves only the mod-squared of the vector and scalar
form factors but no scalar-vector interference term. In the hadronic-rest frame where $\vec{p_k}+\vec{p_{\pi}}=0$, it takes the following form,  
\bea\nonumber
d\Gamma_{SM}= \frac{1}{2m_{\tau}}\times \frac{G_F^2 sin^2 \theta_c}{2}\,V_{us}\,S_{EW}\, \left( m_{\tau}^2-Q^2 \right)\\ \nonumber
\left \{ \left( \frac{2Q^2\,+\,m_{\tau}^2}{3Q^2} \right) 4\,[P(Q^2)]^2\,|F_V|^2\,+ \frac{m_{\tau}^2}{Q^2}\,\frac{(m_K^2 - m_{\pi}^2)^2}{Q^2}|F_S|^2 \right \}dPS^{(3)}~, \\
\eea 
where $S_{EW} = 1.02$~\cite{Amhis:2012bh} is the electroweak correction factor and $P(Q^2)\equiv |\vec{p_k}|$, is the momentum of the kaon in the $K\pi$ rest 
frame, which is a function of the $K\pi$ invariant mass squared $Q^2$, and may be expressed as:
\be
P(Q^2) = \frac{1}{2\sqrt{Q^2}}\sqrt{\left[Q^2-(m_k + m_{\pi})^2\right]\left[Q^2-(m_k - m_{\pi})^2\right]}~.
\ee
After integrating out the neutrino momentum the phase space in the $K\pi$ rest frame is 
\be
dPS^{(3)}= \frac{1}{(4\pi)^3}\frac{ \left( m_{\tau}^2-Q^2 \right)}{m_{\tau}^2}\,|\vec{p_k}|\,\frac{dQ^2}{\sqrt{Q^2}}\,\frac{d\, cos\beta}{2}~,
\ee
where $\beta$ is the direction of the kaon with respect to the tau direction, denoted by $\hat{n}_{\tau}$, viewed from the hadronic rest frame, i.e; 
$cos \beta= \hat{p}_K.\hat{n}_{\tau}$ where $\hat{p}_K=\frac{\vec{p}_K}{|\vec{p}_K|}$~.
Hence, the differential decay rate takes the form,  
\bea \nonumber
 \frac{d\Gamma_{SM}}{d\sqrt{Q^2}}= \frac{G_F^2sin^2\theta_c m^3_{\tau}}{3\times2^5\times\pi^3 Q^2}S_{EW}\left ( 1-\frac{Q^2}{m^2_{\tau}}\right)^2\left (1+\frac{2Q^2}{m^2_{\tau}}\right) \\ 
\times P(Q^2) \left \{ P(Q^2)^2\,|F_V|^2\,+\,\frac{3(m_k^2-m_{\pi}^2)^2}{4Q^2(1+\frac{2Q^2}{m^2_{\tau}})}|F_S|^2 \right \} 
\label{Gtkpnu}~.
\eea

The hadronic current is dominated by many resonances, the vector ones: $K^*(892)$,$K^*(1410)$ and $K^*(1680)$ and the scalars: $K^*_0(800)$ and $K^*_0(1430)$. The form factors can be parameterized in terms of Briet-Wigner forms with energy dependent widths.    
Hence the vector form factor may be written as \cite{Epifanov:2007rf}:
\be
F_V=\frac{1}{1+\beta+\chi}\left[ BW_{K^*(892)}(Q^2) \,+\, \beta BW_{K^*(1410)}(Q^2)\,+\, \chi BW_{K^*(1680)}(Q^2)\right]~,
\label{FV}
\ee
where $\beta$ and $\chi$ are the complex coefficients for the relative contributions of of $K^*(1410)$ and $K^*(1680)$  resonances respectively with respect to the dominant $K^*(892)$ contribution, and $BW_R(s)$ is a relativistic Breit-Wigner function corresponding to $R$ being, $K^*(892)$, $K^*(1410)$ or $K^*(1680)$ for the vector case.

 For each of the resonances, the Breit-Wigner function has the following form:
\be
BW_R(Q^2)=\frac{M_R^2}{Q^2-M_R^2+i\sqrt{Q^2}\Gamma_R(Q^2)}~,
\ee
where 
\be
\Gamma_R(Q^2)=\Gamma_{0R}\frac{M_R^2}{Q^2}\left( \frac{P(Q^2)}{P(M_R^2)}\right)^{(2l+1)}~.
\ee
Here, $\Gamma_R(s)$ is the s-dependent total width of the resonance and $\Gamma_{0R}(s)$ is the resonance width at its peak. The orbital angular momentum $l=1$ if the $K\pi$ system is in a p-wave or a the vector state and $l=0$ for the s-wave or scalar state.\\
The scalar form factor $F_S$ has $K^*_0(800)$ and $K^*_0(1430)$ contributions and has the similar form:
\be
F_S=\varkappa \frac{Q^2}{M^2_{K^*_0(800)}}BW_{K^*_0(800)}(Q^2)\,+\, \gamma\frac{Q^2}{M^2_{K^*_0(1430)}}BW_{K^*_0(1430)}(Q^2)~,
\label{FS}
\ee
where, $\varkappa$ and $\gamma$ are the complex constants that describe the relative contributions of the $K^*_0(800)$ and $K^*_0(1430)$ resonances respectively. 
The Belle collaboration had performed fits to the $K_s\pi^-$ invariant mass spectrum ($Q^2$) distribution and had listed the values of the complex constants $\beta$,$\varkappa$, $\chi$ and $\gamma$ in Ref.~\cite{Epifanov:2007rf}. Their fitted results (as well as those of BaBar reported in \cite{Paramesvaran:2009ec}) demonstrated that a $K^*_{(892)}$ alone is not enough to describe the $K_s \pi$ mass spectrum but rather the distribution shows the clear evidence  for a scalar contribution in the low invariant mass and another component at large $Q^2$.
The fits were best explained with either $K^*(892)+ K^*(1410) + K^*(800)$ or $K^*(892)+ K^*(1430) + K^*(800)$. Therefore we have used these two possibilities for our study and have excluded $ K^*(1680)$ in our study as its inclusion worsen the fit quality in the Belle analysis. We would also like to point out that the Belle fit results are also consistent with a theoretical description using Chiral Perturbation Theory, described in Ref.~\cite{Jamin:2008qg} and that of Ref.~\cite{Moussallam:2007qc} which is based on analycity and $K-\pi$ scattering results. 

\section{Total rate in presence of a new tensor interaction}
\label{Total-Rate}

We now propose an additional tensor contribution  to the amplitude.
We explore if the interaction of the SM with the new tensorial interaction can account for a non-vanishing CP asymmetry in the decay mode $\tau\rightarrow K \pi \nu_{\tau}$. This tensor interaction could arise in various NP models, however, our approach is to study just the effect of this new structure and keep the analysis as model independent as possible. 
Interference of the SM with this tensor amplitude must give a non vanishing CP asymmetry for this particular decay mode, moreover the sign and magnitude the CP asymmetry must be consistent with the observed result. \\
The effective Hamiltonian due to this new operator is written as\\
\be
{\cal H}_{eff}^{NSI}= G^{\prime}\,sin{\theta}_c\,({\bar s}\sigma_{\mu\nu}u)\, ({\bar \nu_{\tau}}\sigma^{\mu\nu}(1+\gamma_5)\tau)~,
\ee
where $G^{\prime}$ is a complex coupling, 
\be
 G^{\prime}\equiv\frac{R_T\,G_F}{\sqrt{2}}~.
\ee
 The new physics amplitude is then given by
\be
{\cal A}_{T}= G^{\prime}\left[ \langle K\pi|{\bar s}\sigma_{\mu\nu}u |0\rangle \right]\,\left[ {\bar u(p_\nu)}\sigma^{\mu\nu}(1+\gamma_5)u(p_\tau) \right]~,
\ee
where $\sigma^{\mu\nu}=\frac{i}{2}\left( \gamma^{\mu}\gamma^{\nu}- \gamma^{\nu} \gamma^{\mu}\right)$ and
the Hadronic current is given by
\be
\langle K(p_k)\pi(p_{\pi})|{\bar s}\sigma_{\mu\nu}u |0\rangle = i\frac{F_T}{(m_K+m_\pi)} \left[ p_k^{\mu}p_{\pi}^{\nu} - p_k^{\nu}p_{\pi}^{\mu}\right]~, 
\ee 
with $F_T$ being the form factor due to the tensor interaction.
In ${\cal A}_{T}$, we include only left handed neutrinos, as the similar term with a right handed neutrino, in the interference of the SM and NP contributions at the decay rate level, will be suppressed by the neutrino mass. We note here that the $A_{T}$ is a special case of the generalised $A^{NSI}$ mentioned in section~\ref{SM-Rate}. More specifically now, the effective amplitude A is now given by,
\bea \nonumber
{\cal A= A_{SM} + |A_T| } \,e^{i\phi}e^{i\delta}\\ 
|{\ca}|^2 =  |{\ca}_{SM}|^2 \,+|{\ca}_T|^2 +
2\,Re\left({\ca}_{SM}{\ca}^{\dag}_T \right)~,
\label{modsq}
\eea 
where
\bea \nonumber
Re({\ca}_{SM}{\ca}^{\dag}_T)\propto 16 m_\tau R_T \,\big[(p_{\nu}.p_{k}).(p_{k}.p_{\pi})-p_{k}^2(p_{\nu}.p_{\pi})-p_{\pi}^2(p_{\nu}.p_{k})
\\
+(p_{\nu}.p_{\pi}).(p_{k}.p_{\pi})\big]\,
|F_V|\,|F_T|cos(\delta_V-\delta_T)
\label{inter}
\eea
and
\bea \nonumber
 |{\ca}_T|^2 \propto 32\,|R_T F_T|^2\,\big[ 2(p_{\nu}.p_{\pi})(p_{\tau}.p_{k})(p_{k}.p_{\pi})+2(p_{\nu}.p_{k})(p_{\tau}.p_{\pi})(p_{k}.p_{\pi})\\ \nonumber
-2\,m_{pi}^2(p_{\tau}.p_{k})(p_{\nu}.p_{k}) \,-\, 2\,m_{k}^2(p_{\tau}.p_{\pi})(p_{\nu}.p_{\pi})\\
+\, \,m_k^2m_{\pi}^2(p_{\tau}.p_{\nu})\,- \,\,(p_{\tau}.p_{\nu})(p_{k}.p_{\pi})\big]~.
\eea
The observables used in this study are obtained after the
integration over the angular variables of the different contributions
in eqn.(\ref{modsq}). In the interference term of the new tensor contribution with the SM
contribution (eqn.(\ref{inter})), the term that arises from the
symmetric ($Q^\mu$) part of the standard current vanishes after this
angular integration, and hence
only the interference of the tensor with the vector
form factor appears.
The full differential decay rate may hence be written as,
\bea \nonumber
d\Gamma & \equiv & \frac{1}{2m_{\tau}}\, \left[ |{\ca}_{SM}|^2 \,+|{\ca}_T|^2 + 2\,Re\left({\ca}_{SM}{\ca}^{\dag}_T \right)      \right]\, dQ^2\\ 
&=& d\Gamma_1 + d\Gamma_2 + d\Gamma_3 
\label{totaleff}
\eea
\bea \nonumber 
{\text where}\,\,\, d\Gamma_1&=& G_F^2\,sin^2\theta_c\,S_{EW}\,\frac{m_{\tau}^3}{64\pi^3}\,\left(\frac{ m_{\tau}^2-Q^2}{ m_{\tau}^2}\right)^2 \,\frac{P(Q^2)}{(Q^2)^{3/2}}\\
&\times & \left\{ P(Q^2)^2 \left( \frac{2Q^2\,+\,m_{\tau}^2}{3m_{\tau}^2} \right)\,|F_V|^2\,+\,\frac{1}{4}\frac{(m_K^2 - m_{\pi}^2)^2}{Q^2}\,|F_S|^2 \right\} \,dQ^2~,
\label{gamma1}
\eea
\bea \nonumber
d\Gamma_{2}&=& G_F^2\,sin^2\theta_c\,S_{EW}\,\frac{m_{\tau}^3}{64\pi^3}\,\left(\frac{ m_{\tau}^2-Q^2}{ m_{\tau}^2}\right)^2 \,\frac{P(Q^2)}{(Q^2)^{3/2}}\\
&\times & \left\{ P(Q^2)^2\,Q^2\left(\frac{Q^2\,+\,2\,m_{\tau}^2}{3m_{\tau}^2}\right)\, R_T^2\,|F_T|^2\right\}  \,dQ^2
\label{gamma2}
\eea
\bea \nonumber
{\text and}\,\,\, d\Gamma_{3}&=& G_F^2\,sin^2\theta_c\,S_{EW}\,\frac{m_{\tau}^3}{64\pi^3}\,\left(\frac{ m_{\tau}^2-Q^2}{ m_{\tau}^2}\right)^2 \,\frac{P(Q^2)}{(Q^2)^{3/2}}\\
&\times & \left\{2\,P(Q^2)^2\,R_T\,|F_V|\,|F_T|\,\frac{Q^2}{m_{\tau}}\,cos\left(\delta_T(Q^2)\,-\,\delta_V(Q^2)\,+\,\phi \right)\right\}\,dQ^2~. 
\label{gamma3}      
\eea
For the conjugate tau decay mode, only the interference term in eq.~(\ref{gamma3}), will differ, having the opposite weak phase $\phi$.

As mentioned above, the interference of the scalar contribution of the
SM and the antisymmetric tensor contribution vanishes. This is similar
to the vanishing of the scalar and the vector interference
contribution in the SM itself. Note that the scalar term is even under
parity, while both vector and tensor are odd under parity, resulting in
only the vector-tensor interference being even under parity and hence surviving 
after the full (parity even) phase space integration. In other words,
once the angular integration is performed, terms that are odd in
cos$\beta$ vanish, however the parity even interference of the vector
and tensor terms contribute to the decay rate even after this integration. 
 
We wish to point out that after completion of this work, we became aware of some earlier papers where tensor interactions had been introduced 
 in semileptonic tau decays, namely Refs.~\cite{GodinaNava:1995jb},
 ~\cite{Delepine:2005tw} and ~\cite{Delepine:2006fv}. We notice
 several differences in our approach and these earlier
 papers. Firstly, Ref.~\cite{GodinaNava:1995jb}, claims that the
 tensor amplitude does not interfere with the SM amplitude, however,
 as shown in our calculations in this section, this is not true. In
 fact, it is exactly this interference that can possibly account for
 the CP violating rate asymmetry. Ref.~\cite{Delepine:2005tw}
 generates a very tiny CP asymmetry ($\approx 10^{-12}$) by second order weak interactions and ~\cite{Delepine:2006fv} has given a numerical estimate of such an asymmetry, but in the context of a SUSY model, unlike our analytical formulae for a generic tensor interaction. Moreover, since this paper preceeded the BaBar rate asymmetry measurement, they have not used the observables to constrain the NP parameters, which we attempt in the following section.    
\section{Estimation of the New Physics parameter from observables}
\label{estimation}
In order to estimate the parameters of the new tensor interaction term, we need to numerically compute the total decay rates for $\tau^+\rightarrow K_s\pi^+ \bar{\nu}$ and  $\tau^{-}\rightarrow K_s\pi^-\nu$, using  eqns.~(\ref{totaleff})-(\ref{gamma3}). We can write the vector and the scalar form factors in terms of the magnitude and the strong phases as, 
\be
F_v= |F_v|e^{i\delta_v(Q^2)},\,\,\text{and}\,\, F_s= |F_s|e^{i\delta_s(Q^2)}.
\ee
Having expressed the form factors in terms of the combinations of Briet Wigner forms of the various resonances, given in eqns.~(\ref{FV}) and (\ref{FS}), the strong phases of the scalar and the vector form factors can be simply extracted from these complex forms. We have used this vector form factor strong phase ($Q^2$ dependent) for our results below. As mentioned in section~\ref{SM-Rate}, Belle proposed two fit models for the $Q^2$ distribution of their data, which had comparable $\chi^2$ values, both having the vector $K^*(892)$ resonance where the data peaks, as well as the scalar $K^*(800)$. In the region around $1.4GeV$, since the data lied much higher than the fitted curve, inclusion of either $K^*(1410)$ or $K^*(1430)$ resulted in a significant improvement in the goodness of fit. Hence, in our analysis, we consider both the possibilities: $K^*(892)+ K^*(1410) + K^*(800)$ and $K^*(892)+ K^*(1430) + K^*(800)$, considering these two cases one at a time and naming them as Case I and Case II respectively. 

Note that a relative orbital angular momentum $l=2$ of the $K\pi$ system would get contribution from a symmetric $2^+$ state, however, since our  NP amplitude consists of an antisymmetric tensor contribution, such a resonant\footnote{A $2^-$ state cannot decay to $K\pi$, for example $K_2(1770)$  does not decay to $K\pi$ but to $K\pi\pi$, as expected by parity conservation.} contribution to the $K\pi$ is not feasible, we take the tensor form factor to be real and hence $\delta$ appearing in eqn.(\ref{Acp}) in presence of the new tensor interaction will be: $\delta= \delta_v$, since, $\delta_T = 0$. A similar tensor form factor had been introduced in the analysis of $K_{e3}$ and $K_{\mu 3}$ data and in fact PDG~\cite{PDG-2012} gives the constraint on the ratio of $|f_T/f_V|$ for these decays. However, with no such existing analysis from experiments nor any theoretical lattice estimates of the tensor hadronic form factor in the $Q^2$ range relevant for the tau decay being considered here, we assume the tensor form factor to be a constant for simplicity, and determine the product of this constant and its coupling strength from the tau decay observables. 

In the presence of this tensorial NSI, the experimentally measured CP asymmetry $A_{CP}^{Exp}$ will be a result of the combination of the CP asymmetry arising from the $K^0$-${\bar K}^0$ mixing and direct CP asymmetry appearing in the tau decay. If the CP asymmetry is a consequence only of the mixing contribution, then this asymmetry depending on the integrated decay times may be expressed as,
\be
A_{CP}^K=\frac{\int_{t_1}^{t_2}\,dt\,\left[\Gamma(K^0(t)\rightarrow \pi\pi)-\Gamma({\bar K^0}(t)\rightarrow \pi\pi)\right]}{\int_{t_1}^{t_2}\,dt\,\left[\Gamma(K^0(t)\rightarrow \pi\pi)+\Gamma({\bar K^0(t)}(t)\rightarrow \pi\pi)\right]}.
\label{AcpK}
\ee
The $K_s$ produced in the tau decay is observed through the ($\pi^+\pi^-$) final state with $m_{\pi\pi}= m_K$. The $\tau^+$ decays to $K^0$ at the time $t_1$ of tau decay ($\tau^-$ decays to ${\bar K^0}$ at $t_1$) and the time difference between the tau decay and the Kaon decay is of the order of the $K_s$ lifetime ($\tau_S$). Hence, in presence of both direct and indirect CP violation, we may express the observed decay rate asymmetries as
 \be
A^{Exp}_{CP}=
\frac{\Gamma(\tau^+\rightarrow K_s\pi^+ \nu)\,\int_{t_1}^{t_2}\,dt\,\Gamma(K^0(t)\rightarrow \pi\pi)-\Gamma(\tau^-\rightarrow K_s\pi^- \nu)\,\int_{t_1}^{t_2}\,dt\,\Gamma({\bar K^0}(t)\rightarrow \pi\pi)}
{\Gamma(\tau^+\rightarrow K_s\pi^+
  \nu)\,\int_{t_1}^{t_2}\,dt\,\Gamma(K^0(t)\rightarrow
  \pi\pi)+\Gamma(\tau^-\rightarrow K_s\pi^-
  \nu)\,\int_{t_1}^{t_2}\,dt\,\Gamma({\bar K^0}(t)\rightarrow
  \pi\pi)},
\label{AcpExp}
\ee 
Interestingly, as shown below, this asymmetry can be factored into  $A_{CP}^K$ and $A_{CP}^{\tau}$ defined in eqs.(\ref{Acp}) and (\ref{AcpK}), where $A_{CP}^{\tau}$ is the CP violation due to the tensorial interaction. Defining, 
\be
\Gamma^{\tau^\pm}\equiv \Gamma(\tau^\pm\rightarrow K_s\pi^\pm \overset{(-)}{\nu_{\tau}}); \,\,\,\,\,\,\,\Gamma^{\overset{(-)}{K^0}}(t)\equiv \Gamma(\overset{(-)}{K^0}(t)\rightarrow \pi\pi)~,
\ee
the difference of the decay rates in the numerator of $A^{Exp}_{CP}$ in eqn.(\ref{AcpExp}) can be written as, 
\bea \nonumber
&&\Gamma(\tau^+\rightarrow K_s\pi^+ \nu)\,\int_{t_1}^{t_2}\,dt\,\Gamma(K^0(t)\rightarrow \pi\pi)-\Gamma(\tau^-\rightarrow K_s\pi^- \nu)\,\int_{t_1}^{t_2}\,dt\,\Gamma({\bar K^0}(t)\rightarrow \pi\pi)\\  \nonumber
&=& \Gamma^{\tau^+}\,\int_{t_1}^{t_2}\,dt\,\Gamma^{K^0}(t)\,-\,\Gamma^{\tau^-}\,\int_{t_1}^{t_2}\,dt\,\Gamma^{\bar K^0}(t)\\  \nonumber
&=& 2\,\left\{\frac{\Gamma^{\tau^+} + \Gamma^{\tau^-}}{2}\,\frac{\int_{t_1}^{t_2}\,dt\,\left[\Gamma^{K^0}(t)-\Gamma^{\bar K^0}(t) \right]}{2}\,+\,\frac{\Gamma^{\tau^+} -\Gamma^{\tau^-}}{2}\,\frac{\int_{t_1}^{t_2}\,dt\,\left[\Gamma^{K^0}(t) + \Gamma^{\bar K^0}(t)) \right]}{2}\right]\\  \nonumber
&=&\frac{1}{2}\,\left\{\Gamma^{\tau^+} +\Gamma^{\tau^-} \right\}\,\int_{t_1}^{t_2}\,dt\,\left(\Gamma^{K^0}(t) +\Gamma^{\bar K^0}(t) \right)\,\left[A_{CP}^K \,+\,A_{CP}^{\tau}  \right]\\  \nonumber
&=&\Gamma_{av}^{\tau^\pm} \,\int_{t_1}^{t_2}\,dt\,\left\{\Gamma^{K^0}(t)+\Gamma^{\bar K^0}(t) \right\}\,\left[ A_{CP}^K \,+\,A_{CP}^{\tau}  \right]~.
\eea
Similarly, the sum, 
\bea \nonumber
&&\Gamma(\tau^+\rightarrow K_s\pi^+ \nu)\,\int_{t_1}^{t_2}\,dt\,\Gamma(K^0(t)\rightarrow \pi\pi)\,+\,\Gamma(\tau^-\rightarrow K_s\pi^- \nu)\,\int_{t_1}^{t_2}\,dt\,\Gamma({\bar K^0}(t)\rightarrow \pi\pi)\\  \nonumber
&=& \Gamma^{\tau^+}\,\int_{t_1}^{t_2}\,dt\,\Gamma^{K^0}(t)\,+\,\Gamma^{\tau^-}\,\int_{t_1}^{t_2}\,dt\,\Gamma^{\bar K^0}(t)\\  \nonumber
&=&\Gamma_{av}^{\tau^\pm} \,\left(1\,+\,A_{CP}^K \,A_{CP}^{\tau}  \right)\,\int_{t_1}^{t_2}\,dt\,\left\{\Gamma^{K^0}(t)+\Gamma^{\bar K^0}(t) \right\}~.
\eea
Therefore the observed asymmetry and the branching ratio for this decay mode can be written as,
\be
A_{CP}^{Exp}=\frac{A_{CP}^{K}\,+\,A_{CP}^{\tau}}{1\,+\,A_{CP}^{\tau}A_{CP}^K}~.
\label{Acpexp}
\ee
and
\be \nonumber
BR=\frac{\Gamma^{\tau^+}+{\bar \Gamma}^{\tau^-}}{2\,\Gamma_{total}}\,\left[ 1\,+\,A_{CP}^{\tau}A_{CP}^K \right]\,\int_{t_1}^{t_2}\,dt\,\left\{\Gamma^{K^0}(t)\,+\,\Gamma^{\bar K^0}(t)\right\}~.
\ee
Using the time dependence of the widths of $K^0$ and ${\bar K^0}$ to $\pi\pi$ for $t_1\leq \tau_S$ and $\tau_S\leq t_2 \leq \tau_L$ 
we can show that,
\be
\int_{t_1}^{t_2}\,dt\,\left(\Gamma^K(t)\,+\Gamma^{\bar K}(t)\right)= \frac{\Gamma(K_s\rightarrow \pi\pi)}{\Gamma_{K_s}}\,\frac{|p|^2+ |q|^2}{4\,{|p|^2 |q|^2}}= BR(K_s\rightarrow \pi\pi)~,
\label{k2pipi}
\ee 
where $p$, $q$ and $\epsilon$ are the standard $K$ mixing parameters and we
ignore terms of order $\epsilon^2$ in the evaluation of the time integrals of the time
dependent decay rates of $K^0$ and $\bar{K}^0$ to $\pi\pi$, as was
done in reference\cite{Grossman:2011zk}. Hence, the
branching ratio for tau decay to $K \pi \nu$ may be written in the
following form, 
\be
BR=\frac{\Gamma_{av}^{\tau^\pm}}{\Gamma_{total}}\,\left[ BR(K_s
  \rightarrow \pi\pi)\right]\,(1\,+\,A_{CP}^{\tau}A_{CP}^K).
\label{finalBR}
\ee
\subsection{Case I}
Here we have used  $K^*(892)+ K^*(1410) + K^*(800)$.
For this case, Figs.(2) and (3) show the dependence of the Form Factors and the strong phase $\delta_V$ respectively, on $\sqrt{Q^2}$.
\begin{figure}[htb]
\label{CaseI-amp}
\begin{minipage}[t]{0.48\textwidth}
\hspace{-0.4cm}
\begin{center}
\includegraphics[width=7cm]{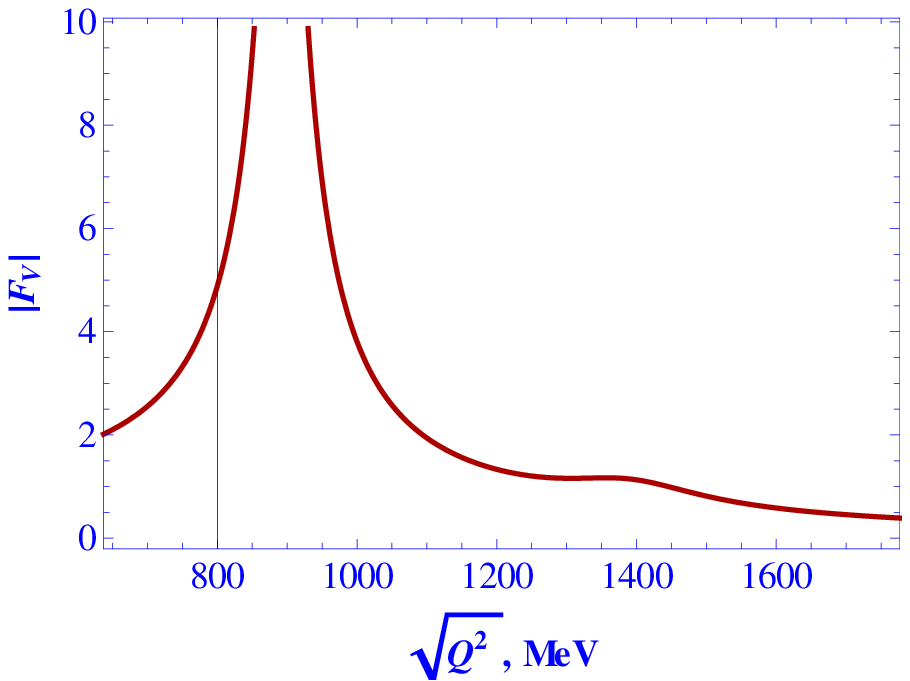}
\end{center}
 \end{minipage}
 \hfill
 \begin{minipage}[t]{0.48\textwidth}
 \hspace{-0.4cm}
 \begin{center}
\includegraphics[width=7cm]{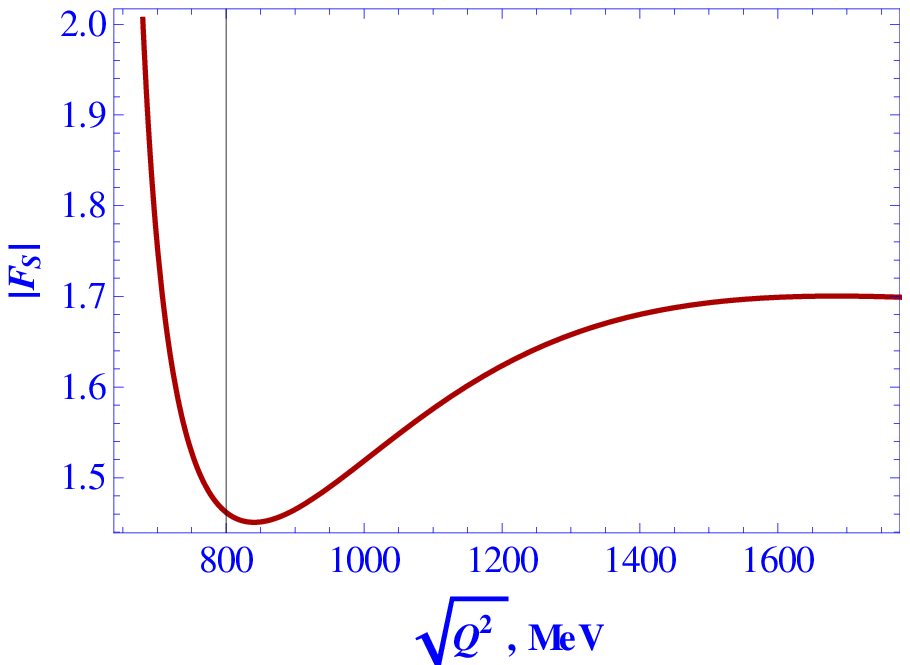}
 \end{center}
 \end{minipage}
\caption{The figures on  the left and the right show the $\sqrt{Q^2}$ dependence of $|F_V|$ and $|F_S|$ respectively, where the parameters appearing in the combination of the Briet Wigner forms for the resonances: $K^*(892)$, $K^*(1410)$ and  $K^*(800)$ were used from the Belle fits for Case I.}
\end{figure}
\begin{figure}[htb]
\label{CaseI-ph}
\hspace{-0.4cm}
\begin{center}
\includegraphics[width=7cm]{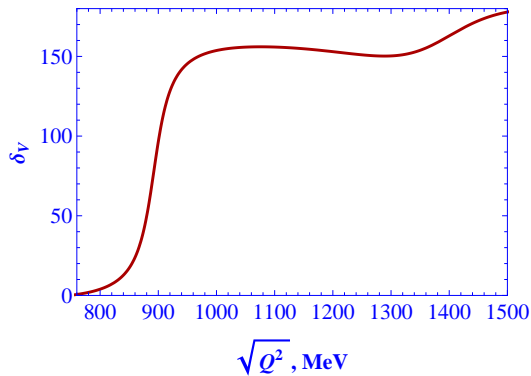}
\end{center}
 \caption{The figure
shows the $\sqrt{Q^2}$ dependence of $\delta_v$ 
as extracted from the complex form with contributions from the combination of the two vector resonances, $K^*(892)$ and $K^*(1410)$, with parameters from Belle fits for Case I.}
 \end{figure}
Substituting these $Q^2$ dependent form factors and the strong phase as well as the values of the masses, decay widths of the various resonances and the branching ratio of the decay mode under consideration
 from PDG~(\cite{PDG-2012},~\cite{PDG-2010}), we compute the $Q^2$ integrated results, $\Gamma^{\tau^\pm}$ for both the decay modes $\tau^{\pm} \rightarrow K_s \pi^{\pm} \overset{(-)}{\nu_{\tau}} $ within the kinematic limits $(m_K+m_\pi)^2$ and $m_\tau^2$. This results in the average effective decay width dependent on the unknown parameters of the new interaction: the product of coupling constant $R_T$ (ratio wrt SM) and the tensor form factor $F_T$ which is assumed to be a constant in this work and the CP violating weak phase $\phi$; numerically the effective widths in MeV are given by, 
\bea  \nonumber
\Gamma^{\tau^\pm} = 8.336\times10^{-12}
\,+\,1.668\times10^{-12}\,(R_T|F_T|)^2 +
2.757\times10^{-12}\,R_T|F_T|\,cos\phi\\
\mp\,R_T|F_T|\,sin(\phi)\,8.52674\times 10^{-13}~. 
\label{GammaI}
\eea In the above equation, the first number is the integrated width
for SM, the second is the width corresponding to the mod-squared of
the tensor contribution, while the last two terms are from the
interference of the SM and tensor parts and hence dependent on the
strong phase ($Q^2$ dependent, which has been integrated out). The
last three terms have been computed in terms of the unknown parameters
of NP, the weak phase $\phi$ and the product of tensor coupling and
form-factor.  From eqn.(\ref{Acp}) we compute the second observable,
the direct CP asymmetry, again in terms of the NP parameters to be,
\be 
A_{CP}^{\tau} = \frac{2\,(R_T|F_T|)\,sin\phi \times 8.527 \times
  10^{-13}}{\Gamma^{\tau^+}+ \Gamma^{\tau^-}}~,
\label{AcpI}
\ee
where, the difference of the widths for $\tau^+$ and $\tau^-$ are
computed, using the difference of $\Gamma_3$ and $\bar{\Gamma_3}$
(after integrating the expressions  in eqn.(\ref{gamma3}) corresponding to those
for $\tau^+$ and $\tau^-$).

Using, the average width, $(\Gamma^{\tau^+}+\Gamma^{\tau^-})/2$
evaluated from eqn. (\ref{GammaI}) and the CP
asymmetry due to direct CP violation 
 calculated in
eqn. (\ref{AcpI}), 
in the expressions for the observed branching ratio and  CP asymmetry
derived earlier (eqns. (\ref{finalBR}) and (\ref{Acpexp})), 
we can find the solutions for the unknowns $R_T|F_T|$ and cos $\phi$. This results in two feasible solutions, displayed in Table~\ref{table-of-r1}.
\begin{table}[htb!]
\begin{tabular}{|c|c|c|c|c|c|}
\hline 
 Sl.No &  $R_T|F_T|$ & cos $\phi$ & $|\frac{T}{SM}|^2$ & $\frac{Int(SM*T)}{SM^2}$ (cos term) &  $\frac{Int(SM*T)}{SM^2}$ (sin term) \\
\hline
(i) & -0.303 & -0.97 & 0.01837 & 0.09721 & 0.00753   \\
\hline
(ii)& -1.945 & -0.99 & 0.75853 & 0.63750 & -0.02809  \\
\hline
\end{tabular}
\caption{Table showing the solutions for the NSI parameters, product
  of ratio of the coupling strength to the SM value and the tensor
  form factor, $R_T|F_T|$, and the cosine of the NP weak phase
  $cos\phi$, allowed by the observables: the branching ratio and the
  CP asymmetry. Columns 4,5 and 6 show the ratio of the contribution
  of the tensor mod-squared term, the interference terms involving the
  cosine of strong and weak phases and that involving the sine of the
  phases, respectively, wrt the SM contribution. The SM part uses the vector and scalar form factors  corresponding to Case I described in the text.}
\label{table-of-r1}
\end{table}
Obviously only the first solution is viable, as the NP contribution has to be much smaller than the SM contribution, since there is no glaring evidence of it, other than the unexpected direct CP violation seen. The smaller magnitude of the tensor mod-squared and interference term relative to the SM contribution, allows the $Q^2$ distribution of SM alone to be reasonably consistent with the Belle data. 
\subsection{Case II}
Here, the combination: $K^*(892)+ K^*(1430) + K^*(800)$ is used.
Figs.(4) and (5) show the dependence of the Form Factors and strong phases, on $\sqrt{Q^2}$ in this case.
\begin{figure}[htb]
\label{CaseII-amp}
\begin{minipage}[t]{0.48\textwidth}
\hspace{-0.4cm}
\begin{center}
\includegraphics[width=7cm]{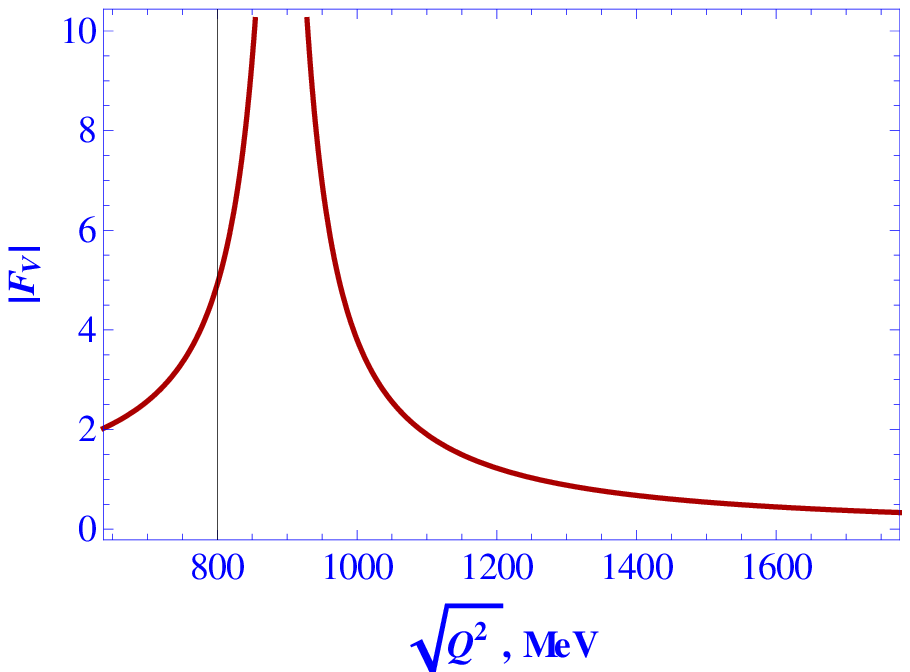}
\end{center}
 \end{minipage}
 \hfill
 \begin{minipage}[t]{0.48\textwidth}
 \hspace{-0.4cm}
 \begin{center}
\includegraphics[width=7cm]{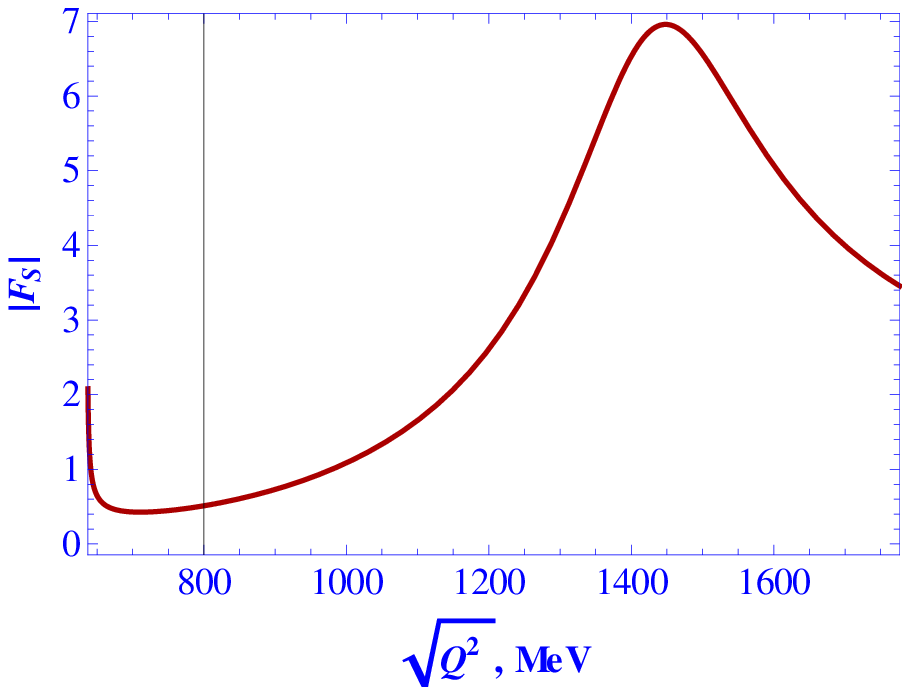}
 \end{center}
 \end{minipage}
\caption{The figure on the left shows the $\sqrt{Q^2}$ dependence of $|F_V|$ and that on the right shows the $\sqrt{Q^2}$ dependence of $|F_S|$, when these form factors include the Briet Wigner contributions from the vector resonance $K^*(892)$ and the scalars: $K^*(1430)$ and $K^*(800)$, as used in the Belle fits for Case II.}
\end{figure}
\begin{figure}[htb]
\label{CaseII-ph}
\hspace{-0.4cm}
\begin{center}
\includegraphics[width=7cm]{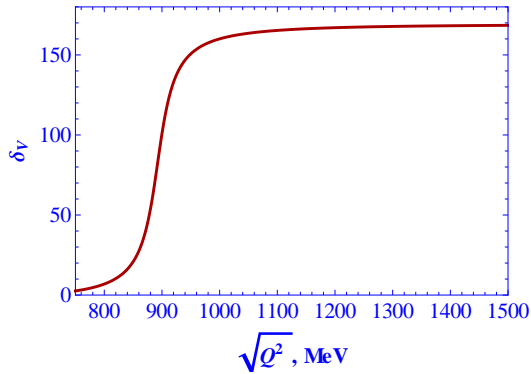}
\end{center}
\caption{The figure
shows the $\sqrt{Q^2}$ dependence of $\delta_v$ 
 corresponding to the vector form factor for Belle fits of Case II}
 \end{figure}
 The complete decay rate is computed to be, 
\bea \nonumber
\Gamma^{\tau^\pm} = 8.294\times10^{-12} \,+\,1.668\times10^{-12}\,(R_T|F_T|)^2 + 2.633\times10^{-12}\,R_T|F_T|\,cos\phi \\
\mp\,5.418\times10^{-13}\,R_T|F_T|\,sin\phi
\eea
From eqn.(\ref{Acp}) we get  
\be
A_{CP}^{\tau} = \frac{2\,R_T|F_T|\,sin\phi \times 5.418 \times 10^{-13}}{\Gamma^{\tau^+}+\Gamma^{\tau^-}}
\label{eqnI}
\ee
Similar to the first case, using the above two equations we get two feasible solutions for $R_T|F_T|$ and cos$\phi$ shown in Table~\ref{table-of-r2} below, where again, only the first solution is meaningful.
\begin{table}[htb!]
\begin{tabular}{|c|c|c|c|c|c|}
\hline 
 Sl.No &  $R_T|F_T|$  & cos $\phi$  & $|\frac{T}{SM}|^2$ & $\frac{Int(SM*T)}{SM^2}$ (cos term) &  $\frac{Int(SM*T)}{SM^2}$ (sin term)\\
\hline
(i) & -0.213 & -0.816 & 0.0091 & 0.05518 & 0.03909  \\
\hline
(ii)& -3.333 & 0.999 & 2.2341 & 1.05703 & 0.04731 \\
\hline
\end{tabular}
\caption{Table showing the the allowed values of the NSI parameters, $R_T|F_T|$ and cos$\phi$, as well as the ratio of the contribution of the tensor mod-squared term wrt the SM contribution, as well as that of the interference contributions, corresponding to the SM form factors for Case II.}
\label{table-of-r2}
\end{table}

In future, once the hadronic form factor for the tensor contribution is estimated theoretically, hopefully from lattice calculations or  a fresh analysis of the larger data sample that may be available\footnote{Belle collaboration, for example has about 3 times larger statistics and has plans to repeat the $\tau\to K_S\pi\nu$ analysis.~\cite{Epf}} is performed by the experimental groups, including the fits with a new tensorial contribution to the amplitude, then, with some handle on the tensor form factor (including its possible $Q^2$ dependence), the coupling strength as well as the weak phase of NP can be pinned down further. Note that the $Q^2$ dependence of the mod-squared of the tensor amplitude is quite different from that of the other terms, which will enable its extraction from data. 

\section{Conclusions:}
\label{conclusion}
CP violation in the quark sector, observed through decays and mixing of $K$ and $B$ mesons, is consistent with its parametrization within the standard model. However, it fails to explain the large baryon asymmetry of the universe and necessitates searches for CP violation beyond the standard  model. Leptonic decays along with semileptonic decays of hadrons  may offer a clean environment for searches of CP violating new physics beyond the standard model.

The recent observation of a CP rate asymmetry $A_{CP}$ by BaBar \cite{BABAR:2011aa} in the tau decay mode $\tau^{\pm} \rightarrow K_s \pi^{\pm} \overset{(-)}{\nu_{\tau}}$ seems to hint at some new physics, with the observed decay rate asymmetry being approximately 2.8 standard deviations away from the standard model predictions of an asymmetry that arises from $K^0-\bar{K}^0$ mixing. The presence of various resonances in the vicinity of the decay hadrons invariant mass, facilitates the availability of strong phases, while complex couplings in a new physics amplitude could provide the weak phase, enabling the possibility of a direct CP asymmetry. A charged scalar contribution can provide a CP violating asymmetry in the angular distribution, but  fails to produce an integrated rate asymmetry. However this is achievable with a generic non-standard tensorial interaction.  We calculated the effective decay rate in the presence of the additional tensor interaction and in fact used the observed branching ratio and CP asymmetry to obtain the parameters of the new physics, the weak phase $\phi$ and product of tensorial coupling and form factor.

\acknowledgements
NS thanks D.~Epifanov for discussions regarding the resonances used in the $Q^2$ distribution fits to data by the Belle collaboration. She thanks Swagato Banerjee and Randall Sobie for communication regarding the $K_S$ lab angle constraints used by the BaBar collaboration in the asymmetry analysis. She is also grateful to  N.~G.~Deshpande and Sandip Pakvasa for their comments.



\begin{thebibliography}{}
\bibitem{Kobayashi:1973fv} 
  M.~Kobayashi and T.~Maskawa,
  Prog.\ Theor.\ Phys.\  {\bf 49}, 652 (1973).
\bibitem{Tsai:1994rc} 
  Y.~S.~Tsai,
  Phys.\ Rev.\ D {\bf 51}, 3172 (1995)
  [hep-ph/9410265];  Y.~-s.~Tsai,
  Phys.\ Lett.\ B {\bf 378}, 272 (1996);  Y.~S.~Tsai,
  Nucl.\ Phys.\ Proc.\ Suppl.\  {\bf 55C}, 293 (1997)
  [hep-ph/9612281].
\bibitem{Bonvicini:2001xz} 
  G.~Bonvicini {\it et al.}  [CLEO Collaboration],
  Phys.\ Rev.\ Lett.\  {\bf 88}, 111803 (2002)
  [hep-ex/0111095].
\bibitem{Bischofberger:2011pw} 
  M.~Bischofberger {\it et al.}  [Belle Collaboration],
  Phys.\ Rev.\ Lett.\  {\bf 107}, 131801 (2011)
  [arXiv:1101.0349 [hep-ex]].
\bibitem{BABAR:2011aa} 
  J.~P.~Lees {\it et al.}  [BABAR Collaboration],
  Phys.\ Rev.\ D {\bf 85}, 031102 (2012)
  [Erratum-ibid.\ D {\bf 85}, 099904 (2012)]
  [arXiv:1109.1527 [hep-ex]].
\bibitem{Bigi:2005ts} 
  I.~I.~Bigi and A.~I.~Sanda,
  Phys.\ Lett.\ B {\bf 625}, 47 (2005)
  [hep-ph/0506037].
\bibitem{Grossman:2011zk} 
  Y.~Grossman and Y.~Nir,
  JHEP {\bf 1204}, 002 (2012)
  [arXiv:1110.3790 [hep-ph]].
\bibitem{delAmoSanchez:2011zza} 
  P.~del Amo Sanchez {\it et al.}  [BABAR Collaboration],
  Phys.\ Rev.\ D {\bf 83}, 071103 (2011)
  [arXiv:1011.5477 [hep-ex]].
\bibitem{Ko:2010ng} 
  B.~R.~Ko {\it et al.}  [Belle Collaboration],
  Phys.\ Rev.\ Lett.\  {\bf 104}, 181602 (2010)
  [arXiv:1001.3202 [hep-ex]].
\bibitem{Mendez:2009aa} 
  H.~Mendez {\it et al.}  [CLEO Collaboration],
  Phys.\ Rev.\ D {\bf 81}, 052013 (2010)
  [arXiv:0906.3198 [hep-ex]].
\bibitem{Link:2001zj} 
  J.~M.~Link {\it et al.}  [FOCUS Collaboration],
  Phys.\ Rev.\ Lett.\  {\bf 88}, 041602 (2002)
  [Erratum-ibid.\  {\bf 88}, 159903 (2002)]
  [hep-ex/0109022].
\bibitem{Bigi:2012km} 
  I.~I.~Bigi,
  arXiv:1204.5817 [hep-ph]; I.~I.~Bigi, 
  arXiv:1210.2968 [hep-ph].
\bibitem{Pich:2009zza} 
  A.~Pich, I.~Boyko, D.~Dedovich and I.~I.~Bigi,
  Int.\ J.\ Mod.\ Phys.\ A {\bf 24S1}, 715 (2009).
\bibitem{PDG-2012}J. Beringer et al. (Particle Data Group), Phys. Rev. D86, 010001 (2012)
\bibitem{Finkemeier:1996dh} 
  M.~Finkemeier and E.~Mirkes,
  Z.\ Phys.\ C {\bf 72}, 619 (1996)
  [hep-ph/9601275]; J.~H.~Kuhn and E.~Mirkes,
  Z.\ Phys.\ C {\bf 56}, 661 (1992)
  [Erratum-ibid.\ C {\bf 67}, 364 (1995)].
\bibitem{Amhis:2012bh} 
  Y.~Amhis {\it et al.}  [Heavy Flavor Averaging Group Collaboration],
  arXiv:1207.1158 [hep-ex].
\bibitem{Epifanov:2007rf} 
  D.~Epifanov {\it et al.}  [Belle Collaboration],
  Phys.\ Lett.\ B {\bf 654}, 65 (2007)
  [arXiv:0706.2231 [hep-ex]].
\bibitem{Paramesvaran:2009ec} 
  S.~Paramesvaran [BaBar Collaboration],
  arXiv:0910.2884 [hep-ex].
\bibitem{Jamin:2008qg} 
  M.~Jamin, A.~Pich and J.~Portoles,
  Phys.\ Lett.\ B {\bf 664}, 78 (2008)
  [arXiv:0803.1786 [hep-ph]].
\bibitem{Moussallam:2007qc} 
  B.~Moussallam,
  Eur.\ Phys.\ J.\ C {\bf 53}, 401 (2008)
  [arXiv:0710.0548 [hep-ph]].
\bibitem{GodinaNava:1995jb} 
  J.~J.~Godina Nava and G.~Lopez Castro,
  Phys.\ Rev.\ D {\bf 52}, 2850 (1995)
  [hep-ph/9506330].
\bibitem{Delepine:2005tw} 
  D.~Delepine, G.~Lopez Castro and L.~-T.~Lopez Lozano,
  Phys.\ Rev.\ D {\bf 72}, 033009 (2005)
  [hep-ph/0503090].
\bibitem{Delepine:2006fv} 
  D.~Delepine, G.~Faisl, S.~Khalil and G.~L.~Castro,
  Phys.\ Rev.\ D {\bf 74}, 056004 (2006)
  [hep-ph/0608008].
\bibitem{PDG-2010}K. Nakamura et al. (Particle Data Group), J. Phys. G 37, 075021 (2010).
\bibitem{Epf}Communication with D. Epifanov. 
\end{thebibliography}
\end{document}